# Parichayana: An Eclipse Plugin for Detecting Exception Handling Anti-Patterns and Code Smells in Java Programs

Ashish Sureka

ABB Corporate Research, Bangalore, India
ashish.sureka@in.abb.com

**Abstract.** Anti-patterns and code-smells are signs in the source code which are not defects (does not prevent the program from functioning and does not cause compile errors) and are rather indicators of deeper and bigger problems. Exception handling is a programming construct designed to handle the occurrence of anomalous or exceptional conditions (that changes the normal flow of program execution). In this paper, we present an Eclipse plug-in (called as Parichayana) for detecting exception handling anti-patterns and code smells in Java programs. Parichayana is capable of automatically detecting several commonly occurring exception handling programming mistakes. We extend the Eclipse IDE and create new menu entries and associated action via the Parichayana plug-in (free and open-source hosted on GitHub). We compare and contrast Parichayana with several code smell detection tools and demonstrate that our tool provides unique capabilities in context to existing tools. We have created an update site and developers can use the Eclipse update manager to install Parichayana from our site. We used Parichyana on several large open-source Java based projects and detected presence of exception handling anti-patterns.

## 1 Features, Hosting, Novelty & Experimental Results

Figure 1 shows a snapshot of Parichayana. The Eclipse plug-in can be used to browse all detected code smells and immediately go to the location of the individual code smells in the Java source code (which is colored) to start refactoring in the Eclipse Java code editor. Markers are the Eclipse mechanism for resource annotations. We use markers to show the user what code smells were detected by the plug-in and where in the source code they can be found. We briefly describe the framework and libraries we use to create and deploy Parichayana. We use Eclipse Luna which includes full support for Java 8 in Java development tools and plug-in development tools. We use the Eclipse Plug-in Development Environment (PDE) which consists of PDE UI (editors, wizards and launchers), API tools and build facilities. We use the Eclipse JDT (Java Development Tools) and the Eclipse AST (Abstract Syntax Tree) libraries to access and read the elements of a Java program. The AST is a detailed tree representation of the Java source code (comparable to the DOM tree model of an XML file). We use the classes in org.eclipse.jdt.core.dom package such as CatchClause for extracting the body of a catch clause.

**Table 1.** List of 14 Anti-Patterns and Code-Smells Detected by Parichayana

|    | CODE | DESCRIPTION |
|----|------|-------------|
| 1  | PSTE | Printing stack-trace and throwing Exception, choose one otherwise it results in multiple log messages (multiple-entries, duplication) |
| 2  | LGTE | Logging and throwing Exception, choose one otherwise it results in multiple log messages (multiple-entries, duplication) |
| 3  | WEPG | Wrapping the exception and passing getMessage() destroys the stack trace of original exception |
| 4  | RRGC | Relying on the result of getCause makes the code fragile, use org.apache.commons.lang.exception.ExceptionUtils.getRootCause(Throwable throwable) |
| 5  | RNHR | Just returns null instead of handling or re-throwing the exception, swallows the exception, losing the information forever |
| 6  | MLLM | Using multi-line log messages causes problems when multiple threads are running in parallel, two log messages may end up spaced-out multiple lines apart in the log file, group together all log messages, regardless of the level |
| 7  | CTGE | Catching generic Exception, catch the specific exception that can be thrown. If swallowing it then a problem but if re-throw then it is OK |
| 8  | LGRN | Log and return null is wrong, instead of returning null, throw the exception, and let the caller deal with it |
| 9  | PSRN | Print stack-trace and return null is wrong, instead of returning null, throw the exception, and let the caller deal with it |
| 10 | THGE | Throws generic Exception, defeats the purpose of using a checked exception, declare the specific checked exceptions that your method can throw |
| 11 | INEE | Ignoring or suppressing InterruptedException with an empty catch-clause is an anti-pattern, empty catch block prevents in determining that an interrupted exception occurred or knowing that the thread was interrupted |
| 12 | LGFT | If this is really a fatal condition then the method should abort and notify the caller of the fatal condition with an appropriate exception rather than only using log.fatal in the catch block |
| 13 | CNPE | NullPointerException is a logical or programming error in the code (result of a bug) and should be eliminated rather than catching. If you anticipate that a null will be returned then explicitly test for it |
| 14 | TNPE | NullPointerException should not be thrown by the program as it is expected that it is thrown by the virtual machine |

Parichayana is a proof-of-concept hosted on GitGub[3] which is a popular web-based hosting service for software development projects. We provide installation instructions and a facility for users to download the software as a single zip-file. Another reason of hosting on GitHub is due to an integrated issue-tracker which makes reporting issues easier by our users (and also GitHub facilitates easier collaboration and extension through pull-requests and forking). Parichayana is easy to install and use through a zip file[4] or update site[5]. We created a Teaser Video[6] and a Screencast[7] for Parichayana.

---

[3] https://github.com/ashishsureka/parichayana

**Table 2.** Exception Handling Anti-Pattern Detected by CHECKSTYLE, FINDBUGS, JEXCRULE, PMD and PARICHAYANA

|    |       | CHECKSTYLE | FINDBUGS | JEXCRULE | PMD | PARICHAYANA |
|----|-------|------------|----------|----------|-----|-------------|
| 1  | PSTE  | -          | -        | -        | -   | •           |
| 2  | LGTE  | -          | -        | -        | -   | •           |
| 3  | WEPG  | -          | -        | •        | •   | •           |
| 4  | RRGC  | -          | -        | -        | -   | •           |
| 5  | RNHR  | -          | •        | -        | -   | •           |
| 6  | MLLM  | -          | -        | -        | -   | •           |
| 7  | CTGE  | •          | •        | -        | •   | •           |
| 8  | LGRN  | -          | -        | -        | -   | •           |
| 9  | PSRN  | -          | -        | -        | -   | •           |
| 10 | THGE  | •          | •        | -        | •   | •           |
| 11 | INEE  | -          | -        | -        | -   | •           |
| 12 | LGFT  | -          | -        | -        | -   | •           |
| 13 | CNPE  | -          | -        | -        | •   | •           |
| 14 | TNPE  | -          | -        | •        | •   | •           |

**Listing 1.1.** Utils.java, (JOOQ - *LGRN*)

```
catch (Exception fatal) {
  log.error("Cannot parse Postgres array: " + rs.getString(index));
  log.error(fatal);
  return null;
}
```

**Listing 1.2.** RemoteJMeterEngineImpl.java, (JMETER - *LGTE*)

```
catch (Exception ex) {
  log.error("rmiregistry needs to be running to start JMeter in server "
      ↪ + "mode\n\t" + ex.toString());
  throw new RemoteException("Cannot start. See server log file.", ex);
}
```

**Listing 1.3.** HttpSSLProtocolSocketFactory.java, (JMETER - *MLLM*)

```
catch (IllegalArgumentException e) {
  log.warn("Could not set protocol list: " + protocolList + ".");
  log.warn("Valid protocols are: " + join(sock.getSupportedProtocols()))
      ↪ ;
}
```

**Listing 1.4.** ContentCryptoMaterial.java, (AWS - *WEPG*)

```
catch (Exception e) {
```

---

[4] https://dl.dropboxusercontent.com/u/48972351/PARICHAYANA.zip
[5] http://www.snpe.rs/in.software.analytics.parichayana.updatesite
[6] https://www.youtube.com/watch?v=MJ_omsGmgvU
[7] http://www.youtube.com/watch?v=QwccDyaxppw

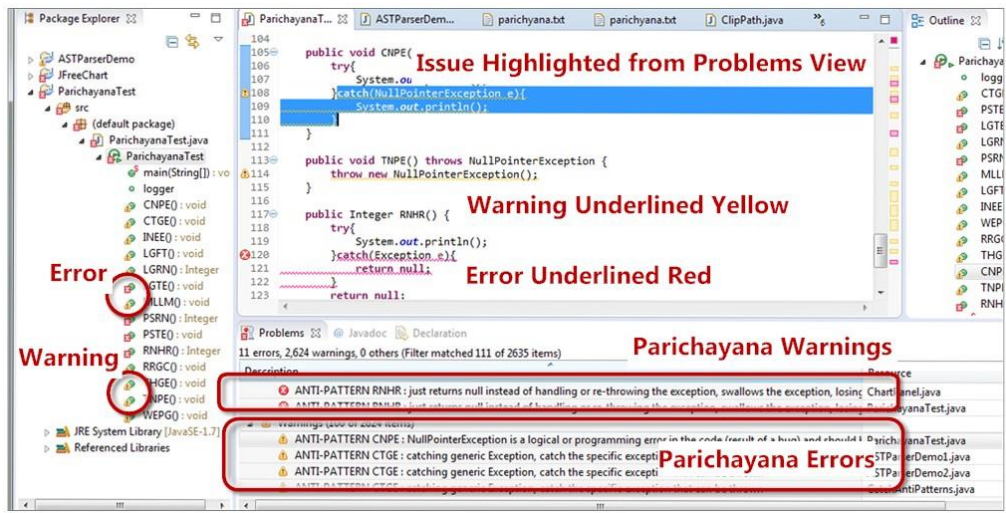

**Fig. 1.** A Screenshot of Parichayana Eclipse Plugin showing the Problems View, List of Anti-Patterns Discovered (Categorized into Errors and Warnings) and the Editor Window showing Problematic Code highlighted and Underlined

```
    throw new AmazonClientException("Error parsing JSON instruction file:
        ↪ " + e.getMessage());
}
```

**Listing 1.5.** Neo4jExceptionTranslator.java, (NEO4J - *RRGC*)

```
catch (IllegalArgumentException iae) {
  if (iae.getCause() != null && iae.getCause() instanceof
       ↪ InvalidEntityTypeException) {
    throw (InvalidEntityTypeException)iae.getCause();
  }
  throw new InvalidDataAccessApiUsageException(iae.getMessage(), iae);
}
```

Table 1 displays the list of 14 anti-patterns[8] and code-smells detected by Parichayana (these are well known gpatterns and not proposed by us – our goal is to build a tool implementing the known ant-patterns). Table 2 shows a feature comparison (on exception-handling anti- patterns) of Parichayana with several widely used and popular code-smell de- tection tool. Table 2 reveals that Parichayana features are novel and unique in context to existing tools. We conduct a series of experiments on several open- source projects (such as Apache JMeter, Apache Tomcat, JFreeChart, Junit, Neo4J and JOOQ) consisting of millions of lines of code and thousands of Java classes. Experimental results demonstrate presence of various exception han- dling anti-patterns and throw light on their intensity. Listings 1.1 − 1.5 shows five different exception handling anti-patterns detected by Parichayana. Listings 1.1 − 1.5 displays the Java class file, name of the open-source project and the anti-pattern detected.

---

[8] https://today.java.net/article/2006/04/04/exception-handling-antipatterns

## 1   Parichayana Installation

Parichayana plug-in can be seamlessly integrated with Eclipse. Parichayana is easy to install and use. Parichayana can be installed similar to other popular third party plugin-ins available for Eclipse (via the Update Manager). The installation processing consisting of Selecting the Install New Software menu item from the Help Menu. Then in the Available Software dialog enter the update site[3] into the work with field. A restart will be required after confirming the unsigned software. Parichayana is easy to install and use through a zip file[4] also. Figure 1 is a snapshot of the Parichayana copyright, version, plug-in details and icon (the command link Help > About Eclipse dialog summarizes information about the installation). We select GPL license so that our code can never be closed-sourced.

## 2   Problems View

Figure 2 shows the snapshot of the Problems View showing problems, errors and warnings logged by Parichayana. Through the Problems View, a user can browse all detected code smells and immediately go to the location of the individual code smells in the Java source code (which is colored) to start refactoring in the Eclipse Java code editor. Double-clicking the icon for a problem, error or warning will automatically open the relevant line of code in the editor for the associated resource. As shown in Figure 2, the problems in the problem view are grouped based on severity. For example, all errors are grouped together. The Problems View shows the name of the exception-handling anti-pattern and its description. Also, multiple filters can be added to the Problems View for enabling or disabling certain problem types.

---

[3]  http://www.snpe.rs/in.software.analytics.parichayana.updatesite
[4]  https://dl.dropboxusercontent.com/u/48972351/PARICHAYANA.zip

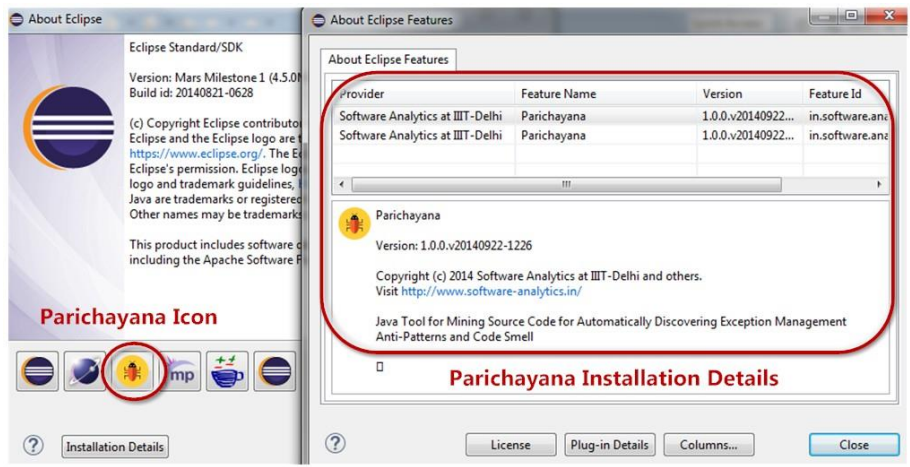

**Fig. 1.** A Snapshot of the Parichayana Copyright, Version, Plug-In Details and Icon

## 3 Editor with Problematic Code

Markers are the Eclipse mechanism for resource annotations. We use markers to show the user what code smells were detected by the Parichayana plug-in and where in the source code they can be found. Problems view shows a list of all errors and warnings (tabular representation: name of the anti-pattern, location, and description) and double-clicking the problem, focuses the Eclipse editor on the marked item. As shown in Figure 3, the marked item in the editor (problematic code) is underlined in either red or yellow color. Errors are shown in red color and warnings are shown in yellow color. Similarly, different icons are displayed for error and warnings. As shown in Figure 3, a tooltip shows-up when the user points the mouse on the marked item or the problematic code.

## 4 Ouput Report

Snapshot (A) in Figure 4 shows the Parichayana output report. Parichayana output report consists of the problematic code and the anti-pattern name and description. The output report consists of only those anti-patterns which are enabled in the preference page. In order to generate the output report, a user needs to right-click on the project and select Open Parichayana Output File (refer to Snapshot (B) in Figure 4). The output report is also generated by selecting Project and then Clean (refer to Snapshot (C) in Figure 4).

## 5 Preference Page

The Parichayana plug-in user is able to make a selection from the set of available code smells, either by making a selection prior to the start of the smell detection process, or by filtering the plug-in results after the detection has taken place. Eclipse's preferences dialog, available from the Window menu, contains a hierarchical list of user preferences. These preference pages are to be used to set preferences for the plug-in, like enabling or disabling detection of a certain type of code smells. Figure 5 shows a snapshot of Parichayana preference page displaying 14 exception handling anti-patterns and a drop-down box from which one can make one of the three selections: Error, Warning and Ignore. A user can search a smaller set of preference titles by using the filter field at the top of the left pane. Preferences in Eclipse are stored as key and value pairs. The key for the preference is an arbitrary String. The value can be a boolean, String, int or another primitive type. The preference support in Eclipse is based on the Preferences class from the org.osgi.service.prefs package. Parichayana also contains a progress monitor to give user the feedback that something is running.

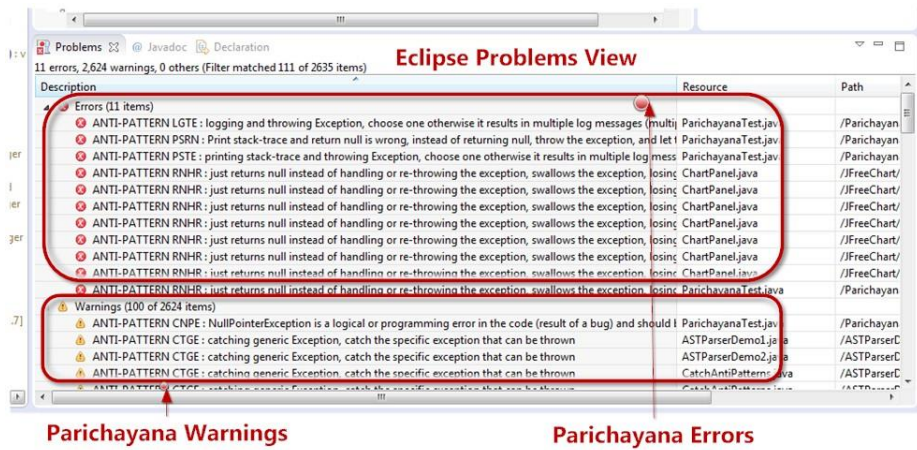

**Fig. 2.** A Snapshot of the Problems View showing Problems, Errors and Warnings logged by Parichayana

## 6 Preference Setting using Regular Expressions

Parichayana provides a per-project basis preference setting. A user can set plug-in preferences on a per-project basis, as opposed to a situation in which settings are made system-wide. Using regular expressions, one can set class and package-level code smell detection. The plug-in presents the user with a user interface as shown in Figure 6, allowing the detection of code smells in selected packages and classes. For example, a caret symbol followed by in\\ will exclude all packages starting with in. The regular expression syntax accepted by Parichayana is the same as used by the java.util.regex API for pattern matching (similar to the Perl programming language). A regular expression can specify complex patterns of character sequences and enables the user to search for patterns in string data by using standardized syntax conventions. Filtering using regular expressions is very useful as one can easily include and exclude files which needs to be processed. We created a Teaser Video[5] and a Screencast[6] for Parichayana.

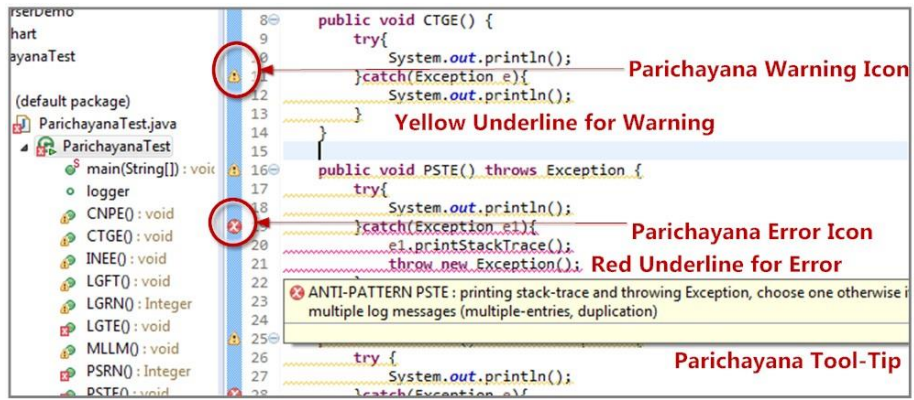

**Fig. 3.** A Snapshot of Eclipse Editor displaying Marked Items (Problematic Code) which are Errors and Warnings. Errors are shown in Red Color and Warnings are shown in Yellow Color

We believe and encourage academic code or software sharing in the interest of improving openness and research reproducibility. We release our exception handling anti-patterns detection tool Parichayana in public domain so that other researchers can validate our scientific claims and use our tool for comparison or benchmarking purposes. We believe our tool has utility and value in the software industry as it helps practitioners in improving code quality and in the spirit of scientific advancement, select GPL license (restrictive license) so that our tool can never be closed-sourced. Some of the close related work are: [1][2][3][4]

### Acknowledgement

We would like to acknowledge Ravindra Naik and Pavan Chittimalli from Tata Research Development and Design Centre (TRDDC) for providing valuable inputs and feedback. We would like to acknowledge Haris Peco for building the user interface.

---

[5] https://www.youtube.com/watch?v=MJ_omsGmgvU
[6] http://www.youtube.com/watch?v=QwccDyaxppw

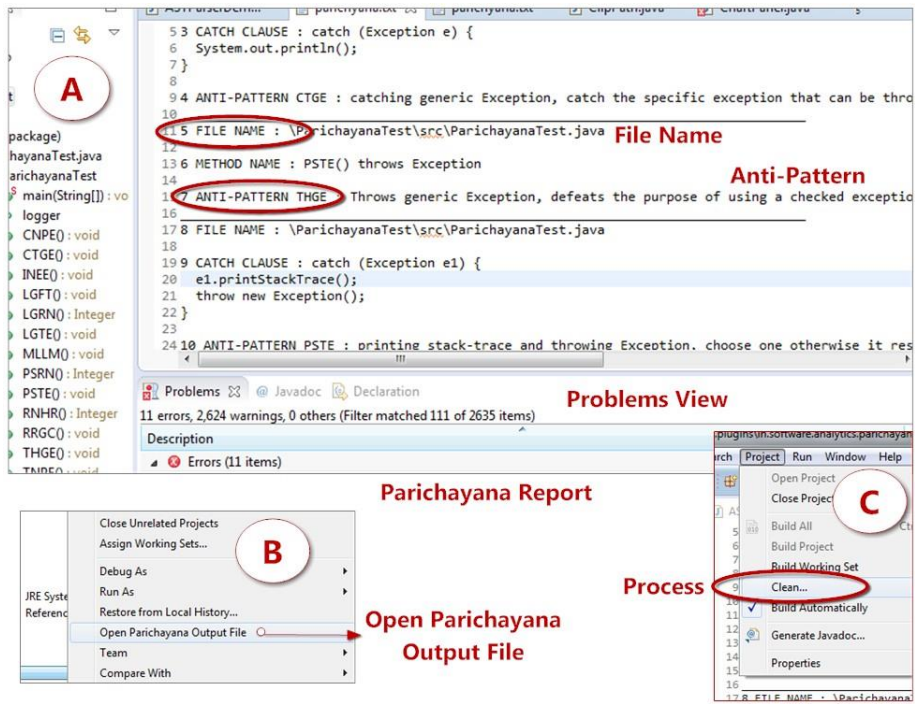

**Fig. 4.** (A) Snapshot of Parichayana Output Report (B) Snapshot of a Contextual Men Item showing how to generate Parichayana Output Report (C) Snapshot of a Menu Item demonstrating how to Execute Parichayana

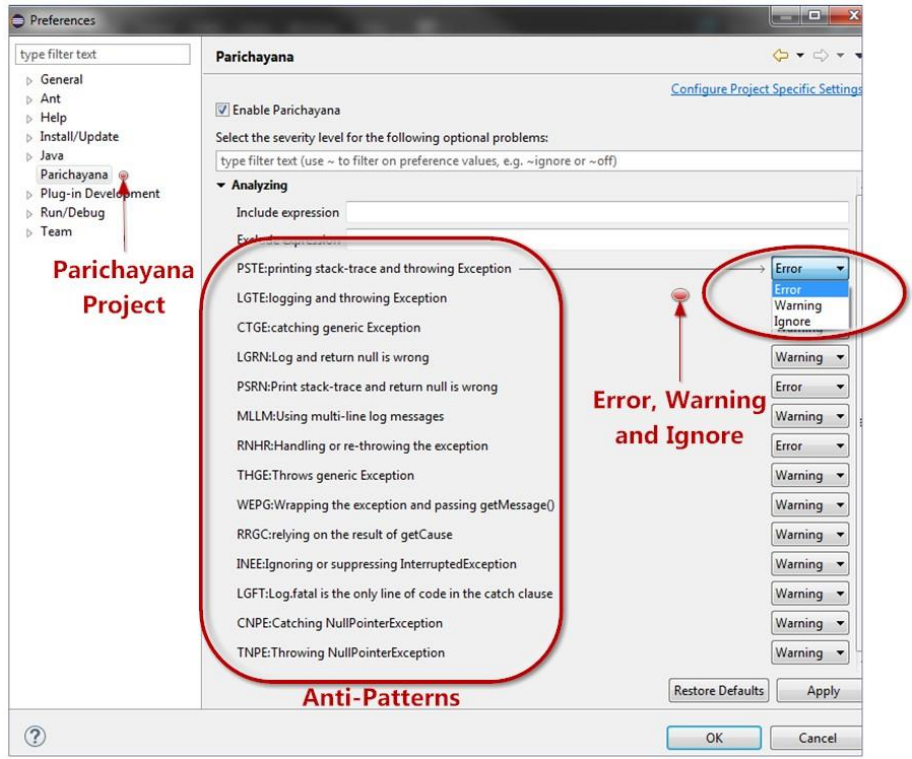

**Fig. 5.** A Snapshot of Parichayana Preference Page displaying 14 Exception Handling Anti-Patterns and a Drop-Down Box from which One can make One of the Three Selections: Error, Warning and Ignore.

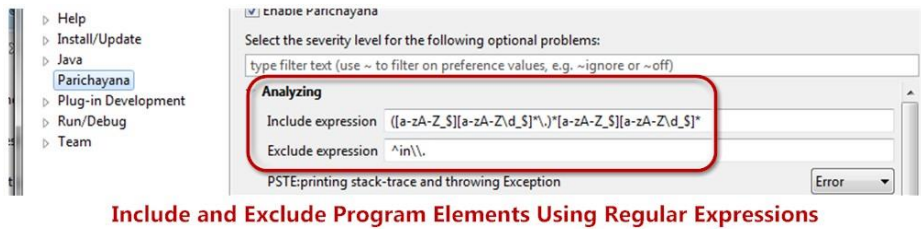

**Fig. 6.** A Snapshot of Parichayana Interface for Filtering Classes and Packages using Regular Expressions